\documentclass{iau}

\usepackage{amsmath}
\usepackage{graphicx}
\usepackage{multirow}
\usepackage[export]{adjustbox}

\begin{document}

\lefttitle{Kerrison \textit{et al.}}
\righttitle{Scintillation: from terrestrial to space weather}

\jnlPage{1}{7}
\jnlDoiYr{2024}
\doival{xxx/xxxxx}
\volno{390}
\aopheadtitle{Proceedings IAU Symposium}
\editors{}

\title{From terrestrial weather to space weather through the history of scintillation}

\author{Emily F. Kerrison$^{1,2,3}$, Ron D. Ekers$^{3}$,  John Morgan$^{4}$ \& Rajan Chhetri$^{4}$}
\affiliation{$^1$Sydney Institute for Astronomy, School of Physics A28, University of Sydney, NSW 2006, Australia
\email{emily.kerrison@sydney.edu.au}}
\affiliation{$^2$ARC Centre of Excellence for All-Sky Astrophysics in 3 Dimensions (ASTRO 3D)}
\affiliation{$^3$CSIRO, Space and Astronomy, PO Box 76, Epping, NSW 1710, Australia}
\affiliation{$^4$CSIRO Space and Astronomy, P.O. Box 1130, Bentley, WA 6102, Australia}

\begin{abstract}
  Recent observations of interplanetary scintillation (IPS) at radio frequencies have proved to be a powerful tool for probing the solar environment from the ground. But how far back does this tradition really extend? Our survey of the literature to date has revealed a long history of scintillating observations, beginning with the oral traditions of Indigenous peoples from around the globe, encompassing the works of the Ancient Greeks and Renaissance scholars, and continuing right through into modern optics, astronomy and space science. We outline here the major steps that humanity has taken along this journey, using scintillation as a tool for predicting first terrestrial, and then space weather without ever having to leave the ground. 

\end{abstract}


\maketitle

\vspace{-5mm}The scintillation of astronomical sources, or the stochastic variation in their phase and amplitude, is so well recognised it has its own nursery rhyme. Yet beyond the ``twinkle twinkle'' of little stars, this phenomenon presents a useful probe of the scattering medium responsible for such variation. In the case of stars visible to the naked eye, this medium is our own atmosphere, while for radio frequency emitting pulsars and the active galactic nuclei (AGN) beyond our own Milky Way, it is typically the interstellar medium or the solar wind (where the effects are referred to as interstellar and interplanetary scintillation, respectively).  It is this last instance, of scattering by the solar wind, that leads directly to predictions for space weather, but to fully appreciate the process of discovery required to reach this level of understanding, it is worthwhile looking back at how both individuals and cultures have used scintillation of various kinds as a type of remote sensing in its simplest form.

\section{Proto-scientific scintillation}

Some of earliest evidence for scintillation-based prediction has survived millenia through the oral traditions of Indigenous peoples in Alaska, South America and Australia. Despite the great length of time from then to now, and the lack of written document, we know that both the Yup'ik of Alaska and the Mocov\'i people of South America were aware of the twinkling of the stars (described in both cultures as ``dancing'' stars), while the Kamilaroi people of south eastern Australia describe them as ``laughing'', and the Wardaman in the far north of the country say they are ``talking'' \citep[][and references therein]{Hamacher-2019}. It is interesting that these disparate cultures, separated by centuries and languages, all draw on the language of human interaction to describe the phenomenon of stellar scintillation. Most interesting for our purposes though, is the fact that, in each of these cultures, these anthropomorphised stars are used as a tool for weather prediction, with stronger scintillation indicative of an impending storm. This has a sound basis in meteorological science, as \cite{sofieva-2013} have shown that stellar scintillation can indeed trace atmospheric turbulence, including (presumably) the strong winds foreshadowing a storm.


Moving from the oral to the written, we see again references to stellar scintillation, and a possible connection with weather prediction, in the writings of the Ancient Greeks. Aristotle wrote of ``the apparent twinkling of the fixed stars and the absence of twinkling in the planets'' in his \emph{De Caelo} some time around 350\,BCE, which is perhaps the first written record of the familiar adage ``stars twinkle, planets don't'' (though in truth this saying is not always accurate, see \citealt{Fuller-2014}). However, given the cosmological nature of this work, he does not tie this statement to weather prediction, instead (incorrectly) rationalising this observation on the basis that starlight has further to travel than planet-light, and so becomes ``weak and wavering'' by virtue of the distance traversed. To find predictive scintillation, we must look at a very different work by a poet Aratus, writing some 80 years later a psuedo-didactic poem called the \emph{Phaenomena}. In this, he talks about the ``darkening disk'' that falls upon stars before rain, and warns his readers to expect a storm ``when the bright light of the stars is dimmed ... and suddenly becomes wavering'' \citep{Kidd_aratus}. This looks very much like the weather prediction seen in the oral traditions of Indigenous peoples above, and suggests that the Greeks too recognised the connection between the strength of stellar scintillation, and the proximity of bad weather.

There are almost certainly other references to stellar scintillation and its use in weather forecasting which are strewn throughout the literature of our collective past. However, even the sample presented here clearly illustrates the longstanding use of twinkling stars as a tool for weather prediction on Earth.

\section{Renaissance and modern science}

Moving from the Ancient world into the Renaissance and finally into what we might reasonably call modern science, scintillation continues to appear sporadically in writings spanning both optics and astronomy. However at this point it is divorced from its connections to weather prediction. Instead, several famous thinkers attempted to explain the physics behind scintillation, but without much more success than Aristotle centuries before them. In the world of optics, da Vinci thought that scintillation was an optical illusion in the eye \citep{Veltman-1986}, an idea which had a curious resurgence in a 1949 \emph{Nature} article \citep{hartridge-1949}. Shortly after da Vinci, Tycho Brahe and Johannes Kepler, both studying the scintillation of supernovae, proclaimed it an intrinsic effect at the source. It was not until Hooke's \emph{Micrographia} (1665), soon followed by Newton's \emph{Opticks} in 1704, that scientists returned to the notion that stellar scintillation might an extrinsic effect caused by the atmosphere millenia after it was used for weather prediction \citep{Monaco-1990}. Yet Newton's focus was on optics, and his explanation dealt only with the effect of aperture size (with larger apertures smoothing out the effects of scintillation). Further discussion of these more recent developments in scintillation can be found in \cite{campbell-1991}, but to return once more to the predictive powers of scintillation we need to switch both wavelengths, and sources.

\section{Scintillation with radio eyes}

From stars to AGN, and optical to radio wavelengths, we see scintillation used once more for remote sensing. Radio frequency scintillation probes three different media, each producing variations on different timescales and discovered in different ways. Ionospheric scintillation probes the upper layers of the atmosphere and manifests as variation on $\sim10$ second timescales, interplanetary scintillation (IPS) probes the solar wind with variation on second timescales, and interstellar scintillation probes the ISM with variation over hours or days \citep[a good summary of these processes can be found in][]{Narayan-1992}. It is the second of these which concerns us here, although all three have their own interesting tales of discovery. 

\begin{figure}[t]\vspace{-1.7cm}
  \centerline{\vbox to 6pc{\hbox to 10pc{}}}
  \raisebox{0.17\height}{\includegraphics[width=0.4\textwidth, trim={0cm 0cm 12cm 0cm},clip, valign=m]{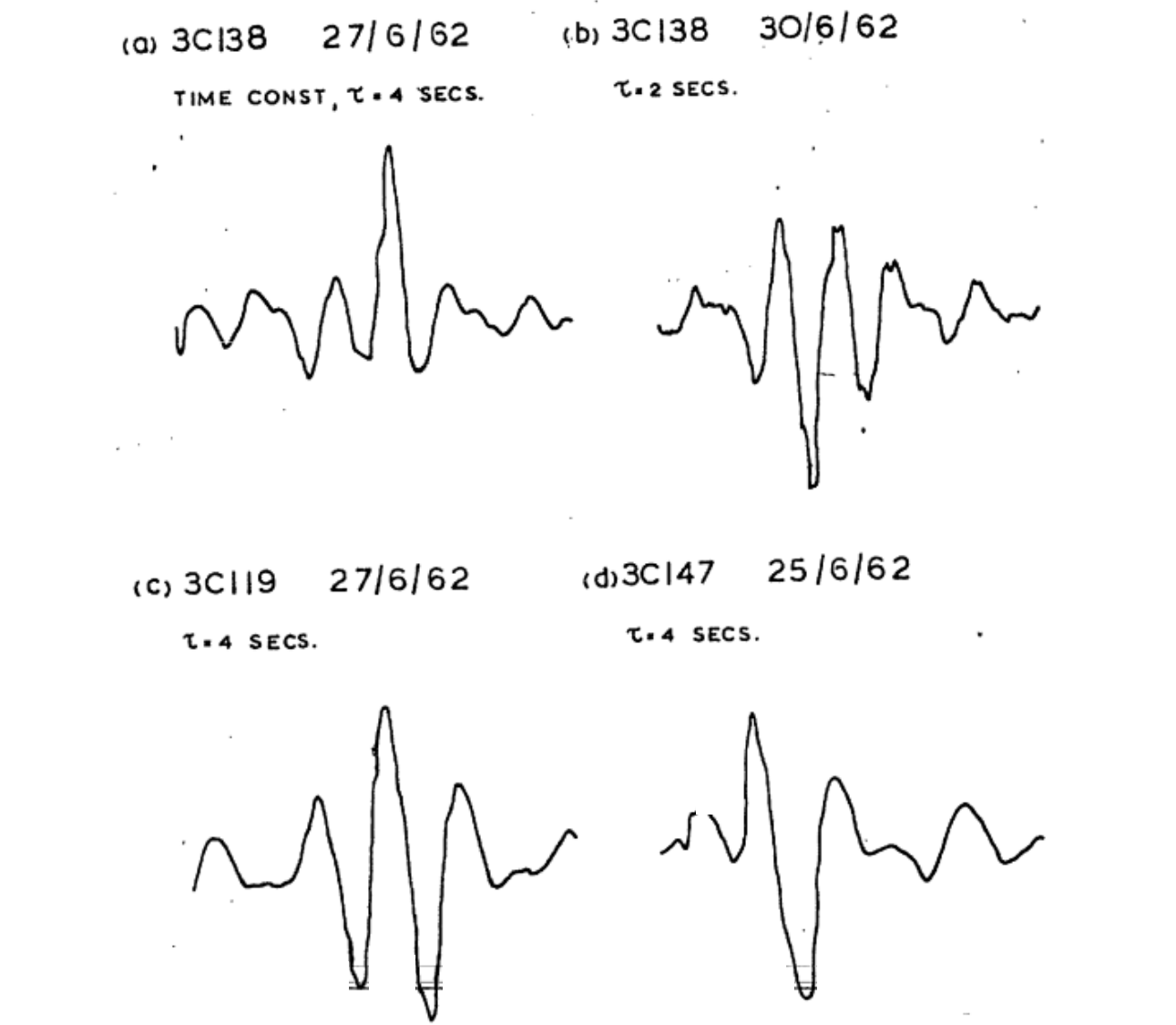}}
  \includegraphics[width=0.56\textwidth, trim={0cm 0cm 0cm 0cm},clip, valign=m]{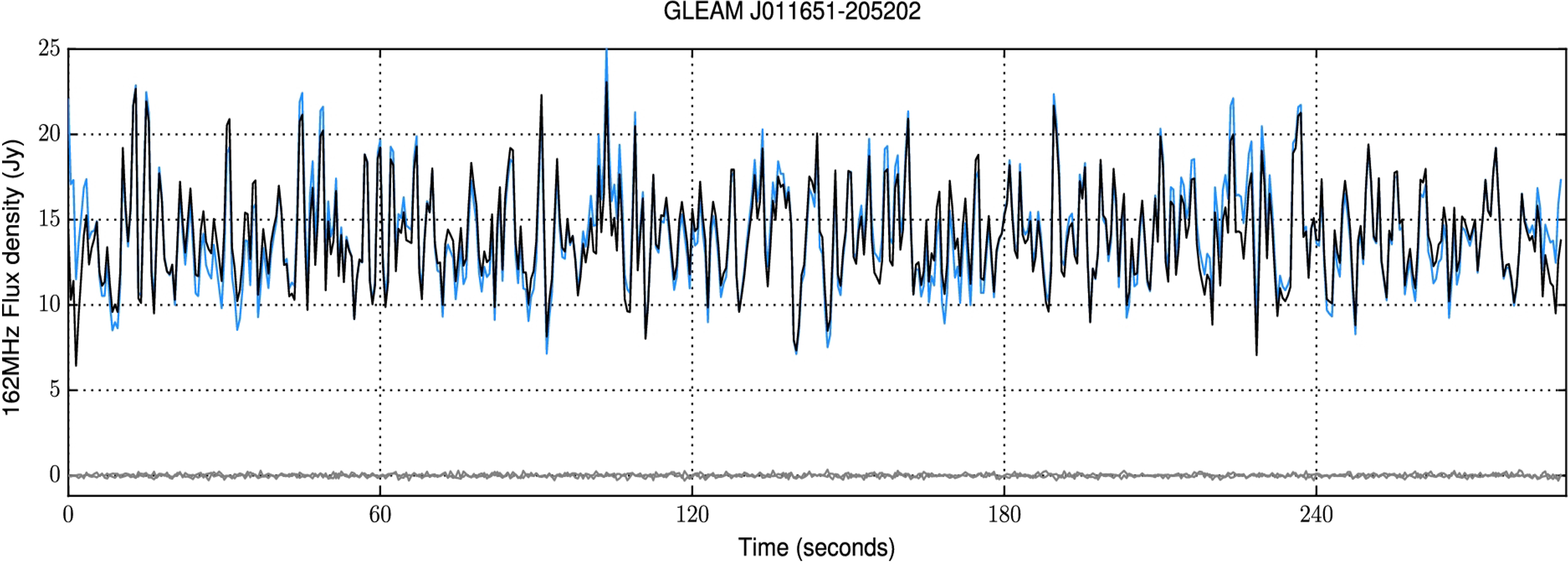}
  \caption{Left: The earliest recorded data exhibiting IPS \citep{Clarke1964IPS}, Right: A typical, modern IPS dataset taken with the Murchison Widefield Array (MWA) telescope with a 0.5 second sampling interval, from \cite{Chhetri2018InterplanetaryFrequencies}.}
  \label{fig:datacompare}
\end{figure}

\vspace{1cm}IPS was first discovered serendipitously by Margaret Clarke, a PhD student studying compact sources. It was published as an appendix to her thesis, in which she correctly identified not only the angular dependence of the phenomenon (that is, it is most visible in sources with angular scales $\theta \lesssim 1 $\,arcsecond), but also the solar corona as the scattering screen \citep{Clarke1964IPS}.  Although her understanding was correct,  it took some months for Tony Hewish, the local Cambridge expert on ionospheric scintillation, to come to terms with this explanation. Once he did, a \emph{Nature} paper soon followed outlining, for the first time, the utility of IPS for making astrophysical measurements of ``quasi-stellar'' sources \citep{Hewish1964InterplanetarySources}. Even so, a few months from discovery to comprehension is a remarkable feat in comparison to the slow growth of knowledge surrounding optical scintillation. Indeed our expertise in this field has grown exponentially, from the Clarke's earliest observations (Figure~\ref{fig:datacompare}, left) to the sophisticated multi-baseline interferometic measurements taken today (Figure~\ref{fig:datacompare}, right).

Today, IPS is a powerful tool for tracking space weather, and there is an entire network of ground-based observatories spanning the globe, dedicated to making IPS measurements. The data are being used as input into heliospheric models to constrain solar wind parameters \citep[e.g.][]{Jackson-2023}. They are used directly to constrain solar wind speeds \citep{Mejia-2015}, and they are used to detect and model coronal mass ejections \citep{Morgan-2023} and stream interaction regions \citep{Waszewski-2023}. Critically, these measurements can be made from the ground, in some cases even using instruments which are shared with astronomical science (e.g. the Murchison Widefield Array \citep{Tingay2013} and Australian SKA Pathfinder, \citep{Hotan2021}), improving the efficiency of space weather tracking and of our scientific instruments more broadly. Recently, space weather IPS measurements have even returned to their roots, being used to study the compact sources behind the scintillating medium \citep{Chhetri2018InterplanetaryFrequencies}.

\section{Scintillation across epochs and frequencies}

This whirlwind tour through the literature captures something of the many ways and many languages in which scintillation has been used, from the oral traditions of Indigenous peoples to the scientific papers of today. There are interesting parallels in the growth of our understanding as to the cause of scintillation, and in its predictive power that span both centuries and continents. It is through this historical appreciation of scintillation that we gain a greater appreciation of scintillation as a tool for space weather, and an example of global science at its best.

\end{document}